\documentclass{article}
\usepackage{makeidx}
\usepackage{multirow}
\usepackage{multicol}
\usepackage[dvipsnames,svgnames,table]{xcolor}
\usepackage{graphicx}
\usepackage{subfigure}
\usepackage{epstopdf}
\usepackage{ulem}
\usepackage{hyperref}
\usepackage{amsmath}
\usepackage{amssymb}
\usepackage{makecell}
\usepackage{algorithm}
\usepackage{algorithmicx}
\usepackage{algpseudocode}
\usepackage{titling}
\usepackage[utf8]{inputenc} 
\usepackage[T1]{fontenc} 
\usepackage{mathrsfs}
\usepackage{mathpazo} 
\usepackage{cite}
\newcommand{\reffig}[1]{Figure \ref{#1}}

\title{Multi-agent navigation based on deep reinforcement learning and traditional pathfinding	algorithm}
\author{Hongda~Qiu}
\usepackage[paperwidth=595pt,paperheight=841pt,top=72pt,right=90pt,bottom=72pt,left=90pt]{geometry}
\makegapedcells
\setcellgapes{5pt}
\makeatletter
	{\par\setlength{\parindent}{#3}
	\setlength{\leftmargin}{#1}       \setlength{\rightmargin}{#2}%
	\advance\linewidth -\leftmargin       \advance\linewidth -\rightmargin%
	\advance\@totalleftmargin\leftmargin  \@setpar{{\@@par}}%
	\parshape 1\@totalleftmargin \linewidth\ignorespaces}{\par}%
\makeatother 
\begin{document}
\maketitle
\begin{abstract}
	We develop a new framework for multi-agent collision avoidance problem.
	The framework combined traditional pathfinding algorithm and reinforcement
	learning. In our approach, the agents learn whether to be navigated or to take
	simple actions to avoid their partners via a deep neural network trained by
	reinforcement learning at each time step. This framework makes it possible for
	agents to arrive terminal points in abstract new scenarios.
	
	In our experiments, we use Unity3D and Tensorflow to build the model and
	environment for our scenarios. We analyze the results and modify the parameters
	to approach a well-behaved strategy for our agents. Our strategy could be
	attached in different environments under different cases, especially when the
	scale is large.
	
\end{abstract}

\section{Introduction}
\begin{figure}[htbp]
\begin{center}
\includegraphics[width=0.6\textwidth]{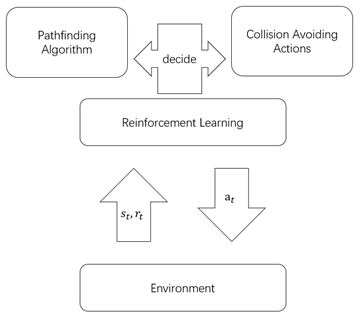}
\caption{An Overview of Our Approach}
\label{img1}
\end{center}
\end{figure}

Multi-agent navigation problem (also known as multi-agent pathfinding
and collision avoidance problem) has been a highly-concerned topic in control,
robotics and optimization. The research of this problem has great potential in engineering applications. One of the major directions is to develop
reliable and high efficient navigating algorithm (or strategy) for agents to
arrive their terminal points without any collisions.\\
In general, a pathfinding and collision avoiding mission requires the
agent(s) to figure out collision-free optimized paths $\gamma$ from the
set-off(s) to the target(s). We can define such a problem rigorously as the following: \\
Let $p$ be an agent, $\alpha$ its set-off, $\beta$ its target and 
$S\subset \mathbb{R}^2$ its motion space with an obstacle set
$\mathcal{B}=\{B_k\vert B_k\subset \mathbb{R}^2,\partial{B_k}\subset C(\mathbb{R}^2)\}$, where $\partial{B_k}$ means the boundary of $B_k$. These boundaries $\partial{B_k}$ can be regarded as closed, continuous loops on Euclidean plane $ \mathbb{R}^2 $. In our work, we further suppose that $\partial{B_k}$ are (piecewise) smooth.\\ 
Suppose the position and velocity of $p$ at time step $t$ are
$\gamma(t)=(x(t),y(t))$, $ \textbf{v}(t) = \dot{\gamma}(x) = (\dot{x}(t), \dot{y}(t)) = (u(t),w(t)) $, respectively, and the maximum time T, the optimization goal of the
problem is:
\begin{equation}
	\min_{\gamma }{\int_0^T{ |  \textbf{v}}_i(t) | dt}
\end{equation}
subject to ${ \gamma}_i(t)\not\in S/ \cup _{B_k\in\mathcal{B}} B_k$, $| { \gamma}_i(t)-\
{ \gamma}_j(t)| \geq d_{safe},$ $
\gamma (0)=s,$ $\gamma (T)=t,\ t\in [0,T]$. Here
$\gamma (t)=({\gamma }_1(t),{\gamma }_2(t))$is the path,
which is second-order differentiable with respect to $t$.\\
Two classes of methods have been developed to solve this problem for the
single-agent case. The first class is usually referred to as continuous methods that employ calculus of variation \cite{Li2017Method}. In those works, the problem is described as an optimal control problem with constraints: given the set of obstacles $\{B_k\}$, the optimization target of some agent $a$ is to minimize the following functional along its path $\gamma $:\\
\begin{equation}
	J(\gamma )=\int_0^TL(t,\gamma (t),\dot{\gamma}(t))dt\ 
\end{equation}
subject to $d({a,\partial B}_k)>d_{safe}$. Here $ L(t,\gamma (t),\dot{\gamma}(t)) $ is an energy functional of the time $t$, the position $ \gamma(t) $ and the velocity $ \dot\gamma(t)$, usually defined by specified circumstances; $ d_{safe} $ is the minimized distance from the $a$ to the bound of $P_k$, and $d_{safe}$, the "safe distance" between $p$ and any obstacles \cite{Li2017Method}. \\
Discretized methods, on the other hand, are also well developed. In early works, researchers discretize the environment of agent(s) into some
grid world or graph and develop efficient algorithms to figure out global optimal
solution for the agent which is allowed to take several discrete actions. Typical discretize algorithm include $A*$ search algorithm \cite{Hart2007A} and Dijkstra`s algorithm \cite{Dial1969Algorithm}. This line of thinking is still efficacious in many fields including engineering, robotics and commercial games, yet it does not work well in complicated cases, such as that with a stochastic environment or large scale. Several methods have been developed to solve this problem. One method related to our work is to use triangular mesh \cite{Debled2002Discrete} instead of grid world to decrease the computational cost \cite{Demyen2007Efficient}. Compared to continuous methods, discretized methods typically require smaller computational cost and have higher
efficiency.\\
Multi-agent navigation \cite{bhattacharya2014multi},\cite{honig2016multi}, \cite{sharon2015conflict}, however, arouses new challenges for us. First of all, in such a system, each agent has to be considered with a group of independent
constraints, which brings large cost for numerical calculation whether we adopt
continuous or discontinuous methods. Secondly, the model of the problem is
NP-hard and the scale of the problem in practice can usually grow extremely large. In practice, every agent need to avoid not only static obstacles but also their partners who themselves are also dynamical. Getting precise solutions numerically via traditional tools is thus expensive and time-costing. \\
Fortunately, some characteristics of the problem offer researchers new
implication. The Markov property of the process of navigation implies the
possibility of using reinforcement learning tools. For each agent, the state of t
$s_t$ is completely determined by $s_{t-1}$ and its decision (or action) at $ t - 1 $,
$a_{t-1}$.
\[
s_0\stackrel{a_0}{\rightarrow }s_1\stackrel{a_1}{\rightarrow }…\stackrel{a_{t-1}}{\rightarrow }s_t
\]
Reinforcement Learning (RL) has immense advantage in dealing with robot
navigation problem. The advantage of reinforcement learning is that we do not
have to precisely "know" exactly the model of the problem; instead, the
algorithm itself could figure it out. Decisions of agents are completely obtained by a parameterized policy which can be optimized (trained) based on training data of the systems. Such path thus circumvent the challenge of defining and verifying a complicated model represented by some group of equations that are usually hard to solve.\\
In \cite{Sugiyama2015Statistical}, the author introduces an example in which the agent (a Khepera robot) is required to explore the environment and avoid collision with obstacles as long as possible. The problem is finally solved by mean-square policy iteration algorithm, where the task of the agent is specified by a reward function.
In the training process, the agent gathers local information of the environment
via 8 distance sensors, and its navigation strategy, also known as behavioral
policy in reinforcement learning, is trained by a self-organizing map \cite{Kohonen1982Self}.\\
Reinforcement learning has also shown its power in solving multi-robot
navigation problems. Some recent works develop high performance navigating policy
for agents via evolutionary reinforcement learning \cite{liu2020mapper} and deep reinforcement learning \cite{Chen2016Decentralized}-\cite{Long2017Towards}, where
decentralized and non-communicative method is used to train their agents. agents gain
local information of the environment and make their decision independently while
updating the same policy. In \cite{Chen2016Decentralized}, researchers study a small scale decentralized
multi-agent collision avoidance problem. In that work a bilateral (two-agent)
value function is defined based on the state of a certain agent and its neighbor.
However, agents in this work are supposed to be omniscient to their surroundings,
which is impossible in real world. As a progress, \cite{Long2017Towards} further studies the
problem in a larger scale, and develops a decentralized framework, in which the
model of the agents is based on a real product. In the training, a group of 20
agents detects surrounding environment through a series of distance sensors and
share the common policy. In their tests, agents are able to finish tasks in many
rich and complex scenarios.\\
Both of these works employ a class of deep reinforcement learning
algorithm known as policy gradient methods (or gradient-based methods) \cite{Schulman2017Proximal}, \cite{Mnih2016Asynchronous}.
Instead of estimating value function, this class of algorithms directly searches for
the optimal parameter set in the parameter set of the policy, allowing
researchers to solve problems with continuous state and action space and
large-scale inputs. \\
There are still some problems remained for us. First of all, deep
reinforcement learning merely guarantees collision-free paths, but not the
shortest ones--these algorithms do not consider the global structure of the maps
(the activity area for the agents), which makes it fairly possible that the final path is not globally optimized. Secondly, training and testing for agents are only done in limited types of scenarios. This leads to weak robustness for variance of environments. Agents may suffer much weaker performance, i.e., getting higher collision rate or longer average path when the environment varies. For instance, a policy trained using small group of agents may not work well for larger group anymore \cite{Long2017Towards}, \cite{godoy2015adaptive}. Retraining the agents for every new environment, on the other hand, can be expensive and typically require extra designing on the algorithm \cite{George_Karimpanal_2018}.\\
The advantages of reinforcement learning and traditional pathfinding
methods inspire us to combine them together---pathfinding problem has been
efficaciously solved by traditional methods, and collision avoidance by
reinforcement learning.\\
In this work, we explore the direction on combining traditional algorithms
and machine learning methods. We propose a framework that merge multi-agent reinforcement learning and classical pathfinding algorithm together. In our framework, the local decision policy on pathfinding and collision avoidance is parameterized and optimized via reinforcement learning using a policy-gradient algorithm\cite{Schulman2017Proximal}, while the decisions of agents include an action that obtained by a determined and global pathfinding algorithm \cite{Hart2007A}. The complete logic of our framework is shown in (\reffig{img1}). Our numerical results show that this framework achieve the following advantages: (1) Strong robustness to variance of environment, including different maps, random starting and terminus points and different density of agents; (2) Higher efficiency of navigation compared to pure learning method \cite{Long2017Towards} on the same path-programming problems; (3) Flexibility. This means that the policy trained from a fundamental scenario can be used in other scenarios without losing too much accuracy; (4) Independent decision. Unlike some previous work using complete centralized framework \cite{jiang2019multi}, while still adopting a centralized training process, our work enables for each agent a decentralized decision process. \\
This paper is organized as follows. In section 2 we introduce some related works in multi-agent reinforcement learning. The model of multi-agent navigation problem and our approach are presented in section 3. The results and the main results of simulations are reported in section 4, and our works are discussed and a conclusion is made in section 5. 

\section{Related Work}
Multi-agent navigation problem has gained a large fragment of attention from researchers, Multi-agent navigation \cite{bhattacharya2014multi},\cite{honig2016multi}, \cite{sharon2015conflict}. Several directions have been explored and many approaches have been developed to attack this problem, including classical algorithm in computer science \cite{bhattacharya2014multi}, constraint optimization \cite{honig2016multi} and conflict based searching \cite{sharon2015conflict}, \cite{9013090}. \\
There has also been increasing focus on solving multi-agent navigation problem
using machine learning and reinforcement learning. The main inspiration of researchers is usually to make the decision policy highly adaptive to unknown or stochastic environments. Some latest progresses have been made via evolutionary artificial neural networks (ANN) \cite{wang1999multi} or reinforcement learning implemented by  evolutionary neural networks \cite{liu2020mapper}. Other works usually aim at coping with very large scale problems in which agents can only gather partial and local information of environment. For instance, in \cite{H2017Guided}, a deep neural work architecture is adopted to cope with the swarm coordinate problem in which agents could only get information from nearby partners. Another related work is \cite{Tai2017Virtual} in which deep reinforcement learning algorithm is used to build a mapless navigator for agents based on locally visual sensors.\\
Multi-agent reinforcement learning (MARL) has gained great interest from
researchers in applied math and computer science. The improvement and comparison in our work are based on a recent study about MARL on $ \mathbb{R}^2 $ \cite{Long2017Towards}, where a policy-gradient method is adopted and several complex scenarios are tested. \\
One of the most important directions that related to our study stands on how to enhance the performance of learning algorithms via attaching the learning algorithm with extra approaches or tools. For instance, imitation learning--introducing expert knowledge to the model \cite{sartoretti2019primal} can improve the robustness of learning and make the agents adaptive to different environments. In another example \cite{jiang2019multi}, randomized decision process is introduced into the framework to decrease the conflicts among agents and improve synergies of the whole system.\\
Some prior works also focus on solving practical
problems via combining MARL with other fields of knowledge. For example, in \cite{9013090}, MARL is combined with conflict based search, enabling the agents to cope with pathfinding problems with sequential subtasks. In another work \cite{Gazzola2016Reinforcement}, Mattia and his group use one-step Q-learning algorithm and wavelet adopted vortex method to build the dynamical model for schools of swimming fish. \\
Merging RL or MARL with other fields of methods is also a direction gaining increasing attention. For instance, Hu proposes the Nash Q-learning algorithm by attaching the Lemke-Houson algorithm \cite{Lemke1964Equilibrium} to the iteration of policy updating to receive a Nash equilibrium for an N-agent game \cite{Hu2003Nash}. In another work \cite{godoy2015adaptive}, an adaptive algorithm that combines MARL and multi-agent game theory is developed to cope with MARL in highly crowded cases.
\section{Approach}
\begin{figure}[htbp]
\begin{center}
\includegraphics[width=0.8\textwidth]{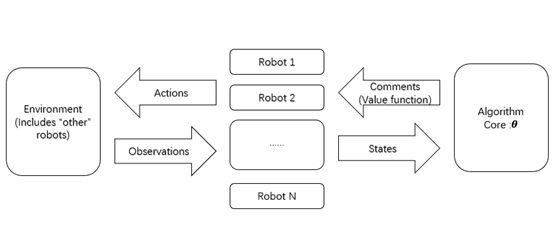}
\caption{Reinforcement Learning}
\label{img2}
\end{center}
\end{figure}

The main idea of our work is to combine traditional pathfinding algorithm and
reinforcement learning. In other words, we employ reinforcement learning to train
the collision-avoidance behavioral strategy for the agents while a pathfinding
algorithm is attached to navigate the agents to their target. The
agents learn to decide whether to be navigated or to avoid their partners at each
time step. Every agent in the scenario acts independently, while training is
centralized (\reffig{img2}) -- all the agents share the same strategy
${\pi }_{\theta }$ and parameter set $\theta $.\\
We begin this section with introducing the model of the problem; then we
introduce the framework of our approach, including the state, action reward
design and algorithm of reinforcement learning.
\subsection{Notations}
Before formal discussions, we introduce some basic notations in this paper.
\begin{itemize}
	\item $\mathbb{R}^d$: $d$-dimensional Euclidean space;
	\item $C(X), C^k(X), C^\infty(X)$: continuous/$k$-degree differentiable/smooth functions on space $ X $ where $ X $ is usually a subspace of some Euclidean space;
	\item $ \Delta X $: the space of distributions on some space $ X $;
	\item $ \mathbb{E}V $: the expectation of some random variable $V$;
	\item $|v|$: norm of vector $v\in\mathbb{R}^d$; 
	\item $\mathcal{S} $: state space of agents in learning;
	\item $\mathcal{A} $: action space of agents in learning;
\end{itemize}
\subsection{The model of the problem}

The model of $N$-agent navigation problem in our work is shown as the
following: Let $i=1,…,N$ be the index of agents. For any agent $i$, let $\alpha_i$ be its starting point and $ \beta_i $ its terminal
point. All agents are running on a motion space $S = [0,l]\times[0,l] \subset \mathbb{R}^2$ with an obstacle set $\mathcal{B}=\{B_k\vert B_k\subset \mathbb{R}^2,\partial B_k\subset C^{\infty }(\mathbb{R}^2)\}$.\\
Suppose that each agent $i$ has a path $\gamma$ and its the position and velocity at time step $t$ is
${\gamma}_i(t)=(x_i,y_i)\in S\subset \mathbb{R}^2$,$\
{ \textbf{v}}_i(t)=(u_i,w_i)\in \left[0,1\right]*[0,2\pi )$
respectively. Furthermore, we note the safe distance between each two agents as $d_{safe}$ and the terminal time as $T_i\leq T$ for each agent $i$. All these variables are listed as in Table\ref{table:variables_of_agents}.\\
\begin{table}[H]
	\centering
	\caption{Variables of Agent}
	\label{table:variables_of_agents}
	\begin{tabular}{c|c}
		\hline
		Notation & Definition\\
		\hline
		$ \gamma_i(t)$ & Position of $i$ at $t$ \\
		$ \textbf{v}_i(t) $ & Velocity of $i$ at $t$ \\
		$ T_i$  & Terminal time of $i$\\
		$d_{safe}$ & The safe distance between agents\\
		\hline
	\end{tabular}
\end{table}
In addition, we assume that:
\begin{itemize}
	\item[a.] $l\gg d_{safe}$;
	\item[b.] Each agent $i$ knows its corresponding $ \alpha_i,\beta_i $ and the structure of the map, but there is no communication among agents.
\end{itemize}
Now it is a fair point to propose the main objective of our approach.\\
First of all, we introduce the goal of policy, that is, the goal of optimization that the optimal policy of agents should achieve.\\
Let $\pi$ be any policy for agents and $\pi*$ be the optimal policy. Consider a discretized partition of the time interval $[0,T]$, $0 < t_1 < t_2 <...< t_{final} = T$. In our work, we take $ t_a = a\Delta t $ where $ \delta t $ notes the time unit for simulation. For simplicity of notation and without loss of generality, we assume that $ T>>1 $ and take $ \delta t = 1 $. Thus, we use can denote any time point by an integer $ t\in\mathbb{Z}_{+} $.\\
As a consequence, we can represent the dynamics of the system as the following:
\begin{equation}
	\label{dynamics}
	\left\{
	\begin{aligned}
		&\gamma_i(t+1) = \gamma_i(t) + v_i(t)\\
		&\gamma_i(0) = \alpha_i, \gamma_i(T_i)=\beta_i\\
		&\gamma_i(t)\notin S/\cup B_k,| { \gamma}_i(t) - { \gamma}_j(t)| \geq d_{safe}\\ 
		&i = 0,1,...,N, t = 0,1,...,T 
	\end{aligned}
	\right.
\end{equation}
In this system, the velocity $ v_i(t) $ is chosen by some policy $\pi$ for each agent $i$ at each time step $t$. \\
Thus, given any set of  $ \alpha_i,\beta_i $, at each time step $t$, a policy $\pi$ can determine a series of paths ${\gamma }_\pi(t;\alpha_i, \beta_i), i = 1,...,N$. To measure how well the policy works, we may compute the average of all the paths, that is,
\begin{equation}
	\label{eq:path_cost}
	S(\pi; \alpha_i,\beta_i, i = 1,...,N) = \frac{1}{N}\sum_{i=1}^NLength({\gamma }_\pi(t;\alpha_i, \beta_i)) = \frac{1}{N}\sum_{i=1}^N\sum_{t=0}^{T_i}|v_i(t)|
\end{equation}
In this work, we wish to optimize $\pi$, i.e., to obtain $\pi*$ among the set of all $\pi$ by apply reinforcement learning. To achieve this, we parameterize $\pi$ with a set of parameters $\theta = (\theta_1,...,\theta_M)\in\mathbb{R}^M$ where $M$ is the dimension. Then we rewrite $\pi$ as $\pi_\theta$.\\
To measure the performance of a policy $\pi_\theta$, we consider a dataset $\mathcal{D}$ of starting points and terminal points, i.e., $ \mathcal{D} = \{ (\alpha_1^m,...,\alpha_N^m;\beta_1^m,...,\beta_N^m), m = 1,...,M\} $. Then we can define a cost function $R$ of $\theta$ by computing the average of $ S(\pi; \alpha_i,\beta_i, i = 1,...,N) $ among all data in $ \mathcal{D} $:
\begin{equation}
	\label{eq:cost_function}
	R(\theta; \mathcal{D}) = \frac{1}{M}\sum_{m = 1}^{M} S(\pi_\theta; \alpha_i,\beta_i, i = 1,...,N) 
\end{equation}
The optimum $\theta*$ is defined as:
\begin{equation}
	\theta*=argmin_{\theta}R(\theta; \mathcal{D})
\end{equation}
The goal of learning is to optimize $ \theta* $ based on some dataset $ \mathcal{D} $. We will discuss our learning framework in the following sections.

\subsection{Multi-Agent Reinforcement Learning}
In this work, we adopt a class of policy gradient method \cite{Schulman2017Proximal},\cite{Mnih2016Asynchronous} to train the policy $\pi_\theta$. This method has shown its capability on solving machine-learning problems with continuous state space in many other works (\cite{Sugiyama2015Statistical}, \cite{Long2017Towards}). We discuss it with more detail in section \ref{sec:policy_gradient_method}. \\
Agents are trained through interaction with the environment (scenarios). In the
simulation, each agent learns whether to find the path to its target, or to take
some simple actions to avoid approaching objects at each time step $t$. The maximum
time step is $T$. The trajectory of each agent is then defined as:
\begin{equation}
	\label{eq:traj_of_each_agent}
	h=\{(s_t,a_t)\vert t=1,2,…,T\}
\end{equation}
In the following part of this section we omit the index i, since the definition
of each agent is exactly the same. The position, velocity, starting point,
terminal point of agent at time step $t$ is then noted as $\gamma(t),v(t),\alpha,\beta$. The terminal time of agent is noted as $T$.\\
Specific model of learning is introduced as the following.

\subsubsection{State}

\begin{figure}[htbp]
\begin{center}
\includegraphics[width=0.4\textwidth]{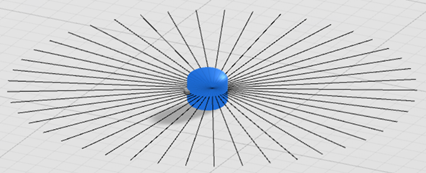}
\caption{The model of robot}
\label{img3}
\end{center}
\end{figure}

In this work, each agent is modeled as disc with radius $ r = 1 $ physical unit,
equipped with 45 range sensors around (\reffig{img3}). The sensors distribute
uniformly around the agent, and each of them return the distance and type (that
is, static or motile) of other objects. The ray distance is noted as $d$.\\
Each agent gains its state at time $t$ $s_t=(o_t)$, where
$o_t=(o_t^j),\ j=1,2,…,45\\ (t=1,2,…,T_i)$. Here $o_t^j=(d_t^j,type_t^j)$
is the information gained by the j the sensor of agent, including distance
$d_t^j\in [0,d]$ and type of perceptible object $type_t^j\in \{0,1\}$ (0 is
referred to static obstacles and 1, other agents) near the agent.

\subsubsection{Action}
The action spaces consist of several actions $\{a_0,a_1,…,a_5\}$, where $a_0$
means to be navigated by some pathfinding method $\mathcal{M}$ in the current time
step, and $a_1,…,a_5$, other simple actions allowed for avoiding collisions. We list all the actions as in \ref{table:list_of_actions}.
\begin{table}[H]
	\centering
	\caption{List of Actions}
	\label{table:list_of_actions}
	\begin{tabular}{c|c}
		\hline
		Notation & Definition\\
		\hline
		$a_0 $ & take action accroding to method $\mathcal{M}$ \\
		$a_1$ & stay at the current point \\
		$a_2$  & move forward: $v=v_0$\\
		$a_3$ & turn left: $v=0,\ \Delta \omega =-\omega_0 $\\
		$a_4$ & turn right: $v=0,\ \Delta \omega =\omega_0 $\\
		$a_5$ & move backward: $v=v_0$\\
		\hline
	\end{tabular}
\end{table}
As a further remark, such construction of action space means that the policy $\pi$ only need to make decision for agents among the six basic actions $a_0,...,a_5$. The specific data of the pathfinding task, i.e., the dataset of $ \mathcal{D} $ and the motion space (in other words, the structure of the map) $ S/\cup_{B_k\in \mathcal{B}} B_k $ are only received by the pathfinding method $A$. In other words, the learning algorithm does not need to consider the specific structure of any map; it only optimize the parameter $\theta$ according to the reward which will be defined in section \ref{sec:reward}.

\subsubsection{Pathfinding Method $\mathcal{M}$}
\begin{figure}[H]
	\begin{center}
		\includegraphics[width=0.6\textwidth]{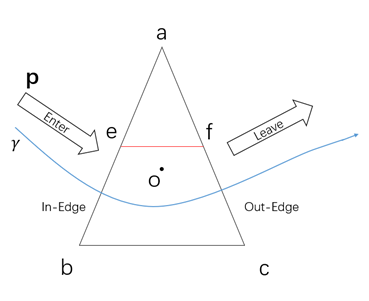}
		\caption{Moving across a triangle: the blue curve $\gamma$ notes the path of $p$ through $\Delta abc$; $p$ enters $\Delta abc$ through edge ab and leave it through edge ac.}
		\label{img20}
	\end{center}
\end{figure}

\begin{figure}[H]
	\begin{center}
		\includegraphics[width=0.6\textwidth]{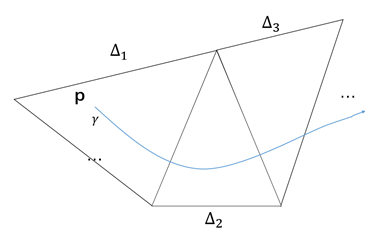}
		\caption{Moving through several triangles}
		\label{img21}
	\end{center}
\end{figure}

Next, we introduce how the method $\mathcal{M}$ works to determine the action $a_0$. We adopt the $A*$ algorithm \cite{Hart2007A}. The map including the obstacle sets $ \mathcal{B} $ is known to the algorithm before we begin to train the agents by reinforcement learning.\\
Given the starting point $\alpha$ and terminus point $\beta$ of a agent $p$, The main idea of $A*$ algorithm is to minimize the following object when $p$ is at point x:
\[
f(x)=g(x)+h(x)
\]
where the cost function $g(x)$ means the cost $p$ has spent on its way from $\alpha$ to
$x$, and the estimation function $h(x)$, an estimate of the cost from $x$ to $t$.\\
In practice we use the Delaunay Triangulation \cite{Debled2002Discrete} to get a navigation mesh on the map and then navigate the agents with $A*$ algorithm. The exact process of $A*$ algorithm could be seen in \cite{Hart2007A}. \\
Before we define $g(x)$ and $h(x)$, we give the definition of cost g and estimation
h of a triangle $\Delta abc$.\\
First of all, we define the "in-edge" and "out-edge" of a triangle
$\Delta abc$ which $p$ passes through. The in-edge is the edge that $p$ go across
when it enters $\Delta abc$; the out-edge, on the opposite, the edge that $p$ go
across when $p$ leaves $\Delta abc$. In \reffig{img20}, suppose that $\gamma $ is the
path of $p$, $ab$ the in-edge and $ac$ the out-edge.\\
The cost $g$ of $\Delta abc$ is defined as the distance between the middle point
of in-edge and the middle point of out-edge. The estimation h is the
straight-line distance from the barycenter $o$ of $\Delta abc$ to the terminus
point $t$of p. In the following picture, for instance, $g=| ef | $ and
$h=\vert ot\vert $.\\
Now we could define $g(x),\ h(x)$ of $p$ in a triangulated map. Suppose
that $p$ is currently at x and has passed through triangles
${\Delta }_1,{\Delta }_2,…,{\Delta }_n$ (\reffig{img21}), where
$x\in {\Delta }_n$, and the cost and estimation of $\Delta $ are $g_i,\ h_i$
respectively, the cost function and estimation function are defined as
$g(x)=\sum_{i=1}^ng_i,\ h(x)=h_n$. Here $s,\ t$ is defined
as above.

\subsubsection{Reward}
\label{sec:reward}
The step reward for each agent is designed as the following:
\[
r_t=r_{navigation}+r_{scenario}+r_{penalty}
\]
Each agent gets a positive reward $ r_{navigation}>0 $ when it decides action
$a_0$. The inspiration of setting so is to encourage the agent to get navigated as frequent as possible so that it can arrive at its target faster. The value of $ \frac{{\pi }_{\theta }(a_t^is_t^i)}{{\pi }_{{\theta }_{old}}(a_t^is_t^i)} $ is set as,
\begin{equation}
	r_{navigation}=\left\{
	\begin{aligned}
		&r_0, &\ if\ decides\ a_0 \\
		&0, &else
	\end{aligned}
	\right.
\end{equation}
The agent also gets a reward $r_{scenario}$ when interacting with the scenario. The idea is straightfoward: if the agent arrives at its target, it would get a positive reward; if it coolides, it would get a negative reward; otherwise, it would get nothing. 
\begin{equation}
	r_{scenario}=\left\{
	\begin{aligned}
		&r_{negative}, &if\ collides\ with\ other\ objects\\
		&r_{positive}, &if\ arrives\ the\ target\\
		&0, &else
	\end{aligned}
	\right.
\end{equation}
Moreover, agent gets a time penalty $r_{penalty}<0$ at each time step. The maximal possible amount of time penalty that an agent can get in one single path shall be smaller than $ T $. To ensure that the agent always gets a positive reward when it manages to arrive at its target, we assume that $ r_{penalty}\cdot T \leq r_{positive}  $.\\
In our experiments we set $r_0=0.00005, r_{negative}=-0.5, r_{positive}=1, r_{penalty}=-0.0001$.

\subsubsection{Neural Network}
\begin{figure}[htbp]
\begin{center}
\includegraphics[width=0.6\textwidth]{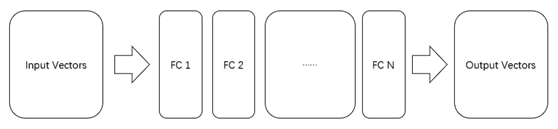}
\caption{The neural network in our approach}
\label{img4}
\end{center}
\end{figure}

Our approach is implemented with a parameter set $\theta \in R^L$ with
scale $L$ for policy $\pi $. Here a linear neural network with $K$ full connection layers, each of $ U = 256 $ units is employed. The input of this network is the observation gained by the sensors, and the output, the action of agent (\reffig{img4}). The scale of parameter set $\theta $ is then $L=256*K$.
\subsubsection{Policy Gradient Method}
\label{sec:policy_gradient_method}
In this work, we adopt a variation of the Policy Gradient Method \cite{Schulman2017Proximal}.\\
Let $ s_t\in\mathcal{S} $ be the state of the agent and $ a_t\in\mathcal{A} $ the action taken by the agent at time step $t$. Let $ V:\mathcal{s}\to\mathbb{R} $ be the value function of learning defined as
\begin{equation}
	\label{eq:value_function}
	V(s) = \mathbb{E}[\sum_{t = 0}^{\infty}\eta^tr_t|s_0 = s]
\end{equation}
where $\eta$ is the discount rate of learning.\\
The policy gradient method can be regarded as a stochastic gradient algorithm aiming at maximizing the following objective,
\begin{equation}
	\label{eq:objective_of_PG}
	L(\theta) = \mathbb{E} \frac{{\pi }_{\theta }(a_t^is_t^i)}{{\pi }_{{\theta }_{old}}(a_t^is_t^i)} \hat{A}_t
\end{equation}
where $ \hat{A}_t $ is the advantage function \cite{Mnih2016Asynchronous} defined as the following:
\begin{equation}
	\label{eq:advantage_function}
	\hat{A}_t = \delta_t + (\eta\lambda)\delta_{t+1} +...+(\eta\lambda)^{T-t+1}\delta_{T-1}
\end{equation}
where $ \delta_t = r_t + \eta V(s_{t+1}) - V(s_t) $. One can think of $\hat{A}_t $ as an analogy of the value function $V$ in classical reinforcement learning framework.\\
Then equation \ref{eq:objective_of_PG} leads to an estimator of policy gradient,
\begin{equation}
	\label{eq:estimator}
	g(\theta) = \mathbb{E} \nabla_\theta \log\pi_{\theta }(a_t|s_t)\hat{A}_t
\end{equation}
where $ \pi_\theta: \mathbb{S}\times\mathcal{A}\to\Delta(\mathcal{A}) $ becomes a randomized policy that determines the distribution of $ a_{t+1} $ according to $ (s_t,a_t) $. 
In this work, the value function of policy $\pi$ is noted as $V(s;\pi_\theta)$. The training process of each agent shares and updates the same parameter set $ \theta $. $U$ is the batch size for data collecting. $\eta ,\
\lambda ,\ \epsilon $ are hyper-parameters of the algorithm. Their values will
be introduced in section \ref{sec:simulation}.
\begin{center}
\vspace{3pt} \noindent
\floatname{algorithm}{Algorithm}
\begin{algorithm*}
	\begin{centering}
	\caption{Proximal Policy Optimization (PPO, use clip surrogate objective) }
    \begin{algorithmic}[1]
        \For{each episode with Max Training Step}
        \For{actor i= 1,2, \ldots , N}
        \State Execute ${\pi }_{{\theta }_{old}}$ for U steps and gain $s_t^i,\ a_t^i,r_t^i$
        respectively;
        \State Calculate advantage estimates $A_1^i…A_t^i$, according to formula:
        \[
        A_t^i=\sum_{p=t}^{T-1}\{{(\eta \lambda )}^{p-t}[r_p^i+\eta V(s_{p+1}^i;\pi_\theta)-V(s_p^i;\pi_\theta)]\}
        \]
        \EndFor
        \State Optimize $L(\theta )=min(u_t^i(\theta )A_t^i,\
        clip(u_t^i(\theta )A_t^i,1-\epsilon ,1+\epsilon )A_t^i)\
        $using the Adam optimizer, where
        $u_t^i(\theta )=\frac{{\pi }_{\theta }(a_t^is_t^i)}{{\pi }_{{\theta }_{old}}(a_t^is_t^i)}$;
        \State ${\theta }_{old}\leftarrow \theta $ (All agents are updated in parallel)
        
        \EndFor
    \end{algorithmic}
    \end{centering}
\end{algorithm*}
\end{center}
\section{Simulation and Results}
\label{sec:simulation}
We begin this section by introducing the training scenario and hyper-parameters
of our experiments. Then we turn to compare our methods quantitively with pure
reinforcement learning methods \cite{Chen2016Decentralized}, \cite{Long2017Towards}. The section ends up with tests on some
special scenarios.

\subsection{Scenarios and Hyper-Parameters}

\begin{figure}[htbp]
\begin{center}
\includegraphics[width=0.4\textwidth]{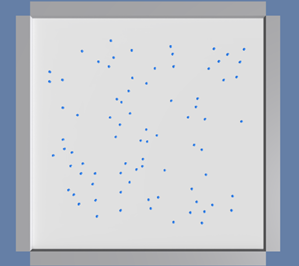}
\caption{The Training Scenario}
\label{img5}
\end{center}
\end{figure}

Our algorithm is implemented in Tensorflow 1.4.0 (Python 3.5.2). The models of
our scenarios and agents are built in Unity3D 2017.4.1f1. The policies are
trained and simulated based on an i7-7500CPU and a NIVIDA GTX 950M GPU.\\
Our basic scenario is built in a square $\alpha$ of size 200 $\times$ 200 (\reffig{img5}).
Initially, $\alpha_i,\beta_i$ are set randomly, and $v_i^0=(0,0)$, with $i=1,2,…,N$. We
set $v_0=1,\ \omega_0 =1,\ T=10000$ and $N=80$. Each agent will be reinitialized once
it arrives its terminus point or its training exceeds the max training time step
$T$. If a collision happens, the involved agents will also be immediately
reinitialized. In the simulation, all the space, angle and velocity in the
experiment adopt the basic units in Unity3D.

The hyper-parameters in table 1.

\begin{figure}[htbp]
\begin{center}
\includegraphics[width=0.4\textwidth]{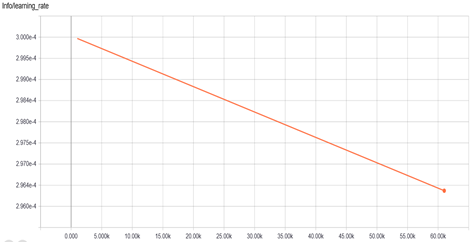}
\caption{The Learning Rate Curve}
\label{fig:learning_rate}
\end{center}
\end{figure}

\begin{table*}[h]
{\centering

\vspace{3pt} \noindent
\begin{tabular}{p{193pt} p{193pt}}
\hline
\parbox{193pt}{\centering 
\textbf{Hyper-Parameters}
} & \parbox{193pt}{\centering 
\textbf{Value}
} \\
\hline
\parbox{193pt}{\centering 
$\eta $
} & \parbox{193pt}{\centering 
0.99
} \\
\parbox{193pt}{\centering 
$ \lambda $

} & \parbox{193pt}{\centering 
0.95
} \\
\parbox{193pt}{\centering 
$ \epsilon $

} & \parbox{193pt}{\centering 
0.1
} \\
\parbox{193pt}{\centering 
Hidden Units
} & \parbox{193pt}{\centering 
256
} \\
\parbox{193pt}{\centering 
Number of Full-Connection Layers
} & \parbox{193pt}{\centering 
4 or 8
} \\
\parbox{193pt}{\centering 
Max Training Step
} & \parbox{193pt}{\centering 
5000000
} \\
\parbox{193pt}{\centering 
Learning Rate
} & \parbox{193pt}{\centering 
0.0003
} \\
\parbox{193pt}{\centering 
Batch Size U
} & \parbox{193pt}{\centering 
2048
} \\
\parbox{193pt}{\centering 
Memory Size
} & \parbox{193pt}{\centering 
2048
} \\
\hline

\end{tabular}
\vspace{2pt}

}
\caption{The hyper-parameter list}
\end{table*}
The learning rate discounts as in \ref{fig:learning_rate}.

\subsection{Comparison: The Baseline Algorithm}
To better demonstrate the performance of our method, we introduce a baseline algorithm \cite{Long2017Towards},\cite{Schulman2017Proximal} and compare the simulation results between our method and this baseline algorithm. \\
The idea of its construction is straightfoward. One can regard this baseline algorithm as a "curtailed" version of our own algorithm in which the action $a_0$ is removed from the action set. In other words, this is an algorithm  that involves nothing more than reinforcement learning. In the following content, we refer to it as "Pure Reinforcement Learning (RL) Method".

\subsection{Results of Simulation}

The result of training and test is shown as the following.

Several measurements are employed to quantitively measure the effectiveness of
our approach in the training scenario and compare it with other methods in
several scenarios. In our tests, a agent begins a trial once it is initialized or
reinitialized. The trial ends if the agent arrives or collides with other
objects. In the tests, we run the model in each scenario for 20000 trials and get
the results.

\begin{enumerate}
	\item Success Rate: In our experiment, a trial of a agent is said to be ``success''
once it arrives its terminus point without colliding others or exceeding max time
step $T$. The success rate is defined as $\frac{Number\ of\ Sucess\ Trials}{Total\
Number\ of\ Trials}$.
	\item Extra Distance Proportion (EDP): Since the starting points and terminus points
are determined randomly, it is unreasonable to measure the circuitousness of the
paths given by our method using average running distance or velocity of the
agents. Hence, we adopt the Extra Distance Proportion (EDP). Given starting point
$a$ and terminus point $b$ of a certain agent, suppose that $l$ is the shortest
distance given by the pathfinding algorithm without considering avoiding its
partner ($l$ could be precalculated before navigation, and in the training
scenario we could obviously get $l=\vert a-b\vert $) and $d$ (obviously
$d\geq l$) is the actual total distance the agent run in practice, the extra
distance proportion $e(a,b)$ is defined as:

\[
e(a,b)=\frac{d-l}{l}
\]

We use $e$ to note the average EDP of all the agents during the test. We could
easily see that the more circuitously the agent runs, the larger $e$ becomes. Note
that EDP are only updated when one agent arrives its terminus point successfully.

	\item Collision Rate: A trial ends and becomes an "accident" if the agent is
involved in a collision. The collision Rate is defined as $\frac{Number\ of\ \
Accidents}{Total\ Number\ of\ Trials}$ during the test, reflecting the
reliability of avoiding collision of the policy.
\end{enumerate}

\begin{figure}[htbp]
	\begin{center}
		\subfigure[]{
			\includegraphics[width=0.5\textwidth]{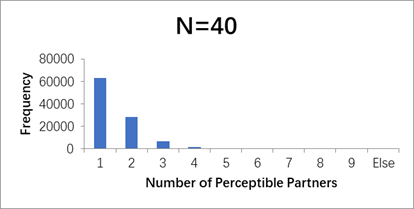}
			
		}
		\subfigure[]{
			\includegraphics[width=0.5\textwidth]{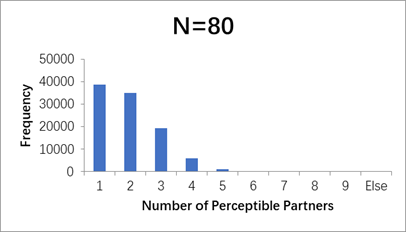}
		}
		\subfigure[]{
			\includegraphics[width=0.5\textwidth]{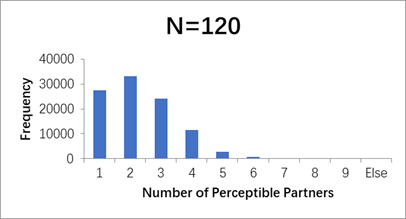}
		}
		\subfigure[]{
			\includegraphics[width=0.5\textwidth]{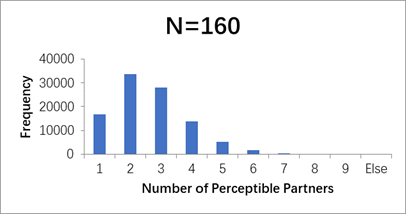}
		}
		\subfigure[]{
			\includegraphics[width=0.5\textwidth]{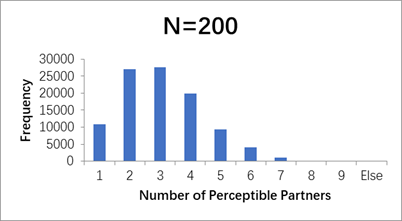}
		}
		\caption{The crowdness when $N=40, 80, 120, 160, 200$}
		\label{fig:frequencies}
	\end{center}
\end{figure}

We train the agents with $N=80$, and test the policy when $N=40, 80, 120, 160, 200$
in the following scenarios. Our method (with 4 or 8 Full -Connection layers) are
compared with the pure reinforcement learning approach in \cite{Chen2016Decentralized}, \cite{Long2017Towards}. In both cases,
the models spend about 4 hours before converging to a stable policy. The
4-FC-layer policy converges after 22000 steps, while the 8-FC-layer case, 66000
steps. The pure RL policy spends 24 hours on training.\\
The learning process is also shown below. \reffig{img7} and \reffig{img8} are screen shots from
Tensorboard. During the training, the Average Reward-Time ($r_t-t$) curve is
adopted here to show the learning effectiveness of our algorithm.
\begin{figure}[htbp]
\begin{center}
\includegraphics[width=0.4\textwidth]{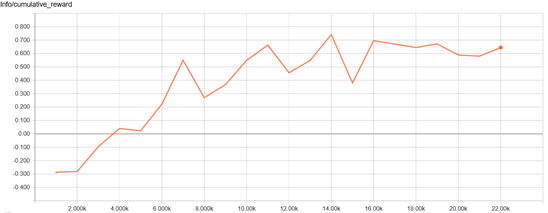}
\caption{The Average Reward-Time Curve of our policy (4 FC layers)}
\label{img7}
\end{center}
\end{figure}
\begin{figure}[htbp]
\begin{center}
\includegraphics[width=0.4\textwidth]{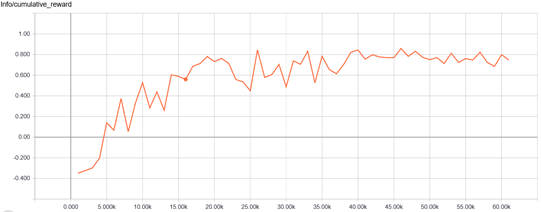}
\caption{The Average Reward-Time Curve of our policy (4 FC layers)}
\label{img8}
\end{center}
\end{figure}
\begin{enumerate}
	\item Basic Scenario: This is our training scenario. During this test we also count the number of perceptible partners of each robot at each time step. We gain a sample of scale 100000 in the case $N = 80, 120, 160, 200$ and draw the histograms of them to show the crowdedness around individual agents. These statistics could partially reflect the variance of observation of agents while $N$ changes (\reffig{fig:frequencies}). Through these histograms we could see that what agents observe when $N=80$ largely differs from that when $N=160$ or $200$. This explains why the performance of our method decrease while $N$ increases, especially when $N$ exceeds 160;
	\item Cross Road Scenario: This scenario is almost the same as the basic scenario. The only thing different is that each agent starts randomly from one of the four edges of $\alpha$ and is supposed to arrive a random point on the opposite edge. For instance, the agent may have starting point $a\in \left[x_1,100\right]\subset S$and terminus point $b\in \left[x_2,-100\right]\subset S$where $x_1,\ x_2\in [-100,100]$. The results are shown in Table 2.
\end{enumerate}
Our policy performances better than pure reinforcement learning methods in the
tests. The agents can arrive at their terminus points with higher accuracy and
lower cost. Meanwhile, our method learns much quicker compared to pure RL method and it does not require the agents to be retrained. It also shows better robustness in different cases, including different density of agents or different maps, especially when the scale of agents increases.\\
Next, we test the policy in different scenarios to show the adaptability for new
scenarios of our approach. The results of these tests are shown in Table 3.
\begin{enumerate}
	\item Four-Wall Scenario;
	\item Random Obstacle Scenario;
	\item Circle Transport scenario. 
\end{enumerate}
In those tests, the agents are capable of coping with unfamiliar scenarios without any further assistance. The value of EDP shows that our policy is able to optimize the paths of agents with lower computational cost, comparing to the pure RL method.
\begin{center}
	\begin{table*}[htbp]
		\vspace{3pt} \noindent
		\begin{tabular}{p{18pt}p{135pt}p{62pt}p{84pt}p{77pt}}
			\hline
			\parbox{18pt}{\centering 
				\textbf{N}
			} & \parbox{135pt}{\centering 
				\textbf{Method}
			} & \parbox{62pt}{\centering 
				\textbf{Success Rate}
			} & \parbox{84pt}{\centering 
				\textbf{EDP (mean/std)}
			} & \parbox{77pt}{\centering 
				\textbf{Collision Rate}
			} \\
			\hline
			\parbox{18pt}{\centering 
				\textbf{40}
			} & \parbox{135pt}{\centering 
				Pure RL Method
				
				Our Method (4 FC layers)
				
				Our Method (8 FC layers)
			} & \parbox{62pt}{\centering 
				0.983
				
				\textbf{0.984}
				
				0.983
			} & \parbox{84pt}{\centering 
				0.573/0.236
				
				\textbf{0.516/0.218}
				
				0.545/0.246
			} & \parbox{77pt}{\centering 
				0.016
				
				\textbf{0.015}
				
				0.016
			} \\
			\hline
			\parbox{18pt}{\centering 
				\textbf{80}
			} & \parbox{135pt}{\centering 
				Pure RL Method
				
				Our Method (4 FC layers)
				
				Our Method (8 FC layers)
			} & \parbox{62pt}{\centering 
				0.963
				
				\textbf{0.971}
				
				0.961
			} & \parbox{84pt}{\centering 
				0.877/0.3252
				
				\textbf{0.652/0.281}
				
				0.701/0.2933
			} & \parbox{77pt}{\centering 
				0.036
				
				\textbf{0.028}
				
				0.038
			} \\
			\hline
			\parbox{18pt}{\centering 
				\textbf{120}
			} & \parbox{135pt}{\centering 
				Pure RL Method
				
				Our Method (4 FC layers)
				
				Our Method (8 FC layers)
			} & \parbox{62pt}{\centering 
				0.898
				
				\textbf{0951}
				
				0.921
			} & \parbox{84pt}{\centering 
				0.987/0.477
				
				\textbf{0.854/0.403}
				
				0.866/0.425
			} & \parbox{77pt}{\centering 
				0.093
				
				\textbf{0.047}
				
				0.077
			} \\
			\hline
			\parbox{18pt}{\centering 
				\textbf{160}
			} & \parbox{135pt}{\centering 
				Pure RL Method
				
				Our Method (4 FC layers)
				
				Our Method (8 FC layers)
			} & \parbox{62pt}{\centering 
				0.732
				
				\textbf{0.862}
				
				0.860
			} & \parbox{84pt}{\centering 
				1.132/0.532
				
				1.054/\textbf{0.456}
				
				\textbf{0.991/}0.481
			} & \parbox{77pt}{\centering 
				0.138
				
				\textbf{0.136}
				
				0.137
			} \\
			\hline
			\parbox{18pt}{\centering 
				\textbf{200}
			} & \parbox{135pt}{\centering 
				Pure RL Method
				
				Our Method (4 FC layers)
				
				Our Method (8 FC layers)
			} & \parbox{62pt}{\centering 
				0.532
				
				\textbf{0.831}
				
				0.783
			} & \parbox{84pt}{\centering 
				1.232/0.738
				
				1.126/0.653
				
				\textbf{1.103/0.689}
			} & \parbox{77pt}{\centering 
				0.249
				
				\textbf{0.167}
				
				0.211
			} \\
			\hline
		\end{tabular}
		\vspace{2pt}
		
		\caption{The results of tests in the basic scenario}
	\end{table*}
\end{center}
\begin{figure}[htbp]
\begin{center}
\includegraphics[width=0.4\textwidth]{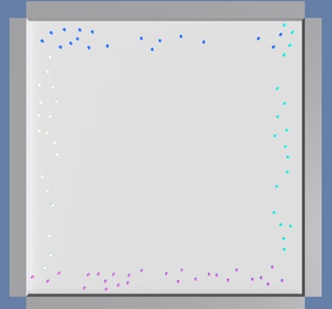}
\caption{Cross Road Scenario}
\label{img14}
\end{center}
\end{figure}
%
\begin{center}
\begin{table*}[H]
\vspace{3pt} \noindent
\begin{tabular}{p{18pt}p{135pt}p{62pt}p{84pt}p{77pt}}
\hline
\parbox{18pt}{\centering 
\textbf{N}
} & \parbox{135pt}{\centering 
\textbf{Method}
} & \parbox{62pt}{\centering 
\textbf{Success Rate}
} & \parbox{84pt}{\centering 
\textbf{EDP}
} & \parbox{77pt}{\centering 
\textbf{Collision Rate}
} \\
\hline
\parbox{18pt}{\centering 
\textbf{40}
} & \parbox{135pt}{\centering 
Pure RL Method

Our Method (4 FC layers)

Our Method (8 FC layers)
} & \parbox{62pt}{\centering 
\textbf{0.999}

\textbf{0.999}

\textbf{0.999}
} & \parbox{84pt}{\centering 
0.676/0.276

\textbf{0.567/0.245}

0.595/0.265
} & \parbox{77pt}{\centering 
\textbf{$<$0.001}

\textbf{$<$0.001}

\textbf{$<$0.001}
} \\
\hline
\parbox{18pt}{\centering 
\textbf{80}
} & \parbox{135pt}{\centering 
Pure RL Method

Our Method (4 FC layers)

Our Method (8 FC layers)
} & \parbox{62pt}{\centering 
0.921

\textbf{0.999}

\textbf{0.999}
} & \parbox{84pt}{\centering 
1.021/0.323

0.759/0.293

\textbf{0.756/0.301}
} & \parbox{77pt}{\centering 
0.053

\textbf{$<$0.001}

\textbf{$<$0.001}
} \\
\hline
\parbox{18pt}{\centering 
\textbf{120}
} & \parbox{135pt}{\centering 
Pure RL Method

Our Method (4 FC layers)

Our Method (8 FC layers)
} & \parbox{62pt}{\centering 
0.865

\textbf{0.992}

0.973
} & \parbox{84pt}{\centering 
1.352/0.501

\textbf{0.861/0.412}

0.959/0.454
} & \parbox{77pt}{\centering 
0.082

\textbf{0.007}

0.025
} \\
\hline
\parbox{18pt}{\centering 
\textbf{160}
} & \parbox{135pt}{\centering 
Pure RL Method

Our Method (4 FC layers)

Our Method (8 FC layers)
} & \parbox{62pt}{\centering 
0.742

\textbf{0.945}

0.923
} & \parbox{84pt}{\centering 
1.932/0.572

\textbf{1.030/0.478}

1.054/0.492
} & \parbox{77pt}{\centering 
0.142

\textbf{0.054}

0.071
} \\
\hline
\parbox{18pt}{\centering 
\textbf{200}
} & \parbox{135pt}{\centering 
Pure RL Method

Our Method (4 FC layers)

Our Method (8 FC layers)
} & \parbox{62pt}{\centering 
0.632

\textbf{0.914}

0.889
} & \parbox{84pt}{\centering 
2.214/0.784

\textbf{1.171/0.677}

1.170/0.709
} & \parbox{77pt}{\centering 
0.232

\textbf{0.085}

0.109
} \\
\hline
\end{tabular}
\vspace{2pt}

\caption{The results of tests in the cross-road scenario}
\end{table*}
\end{center}
\begin{figure}[htbp]
	\begin{center}
		\includegraphics[width=0.4\textwidth]{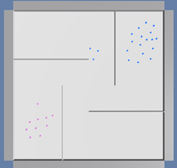}
		\caption{The Four-Wall(FW) Scenario: In this scenario, the blue agents start
			from $\{(x,y)| x\in 80,100, y\in [80,100]\}$ and are supposed to run
			to $\{(x,y)|x\in -80,-100, y\in [-80,-100]\}$; the pink agents do the
			opposite. Here we set $N=30$.}
		\label{img16}
	\end{center}
\end{figure}
\begin{figure}[htbp]
	\begin{center}
		\includegraphics[width=0.4\textwidth]{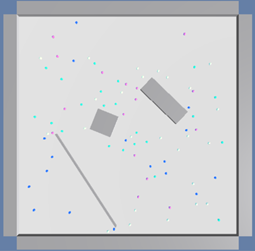}
		\caption{The Random Obstacle (RO) Scenario: This scenario contains several rectangle
			obstacles of random size. The starting and terminus points of agents are randomly
			set just like that in the cross road scenario. Here we set N=80, 120.}
		\label{img17}
	\end{center}
\end{figure}
\begin{figure}[htbp]
	\begin{center}
        \includegraphics[width=0.4\textwidth]{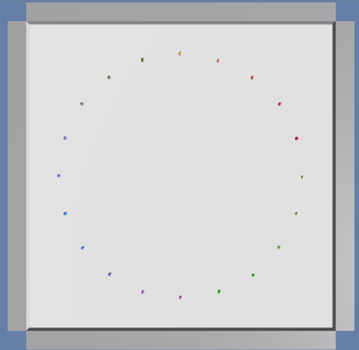}
        \caption{The Circle Transport (CT) Scenario: In this scenario, 20 agents are put uniformly on a circle of radius=80. Each of them are supposed to arrive its antipodal point. This is a difficult mission in multi-agent navigation, but the agents successfully finish the mission under the navigation.}
        \label{img18}
    \end{center}
\end{figure}
\begin{center}
\end{center}
\begin{center}
\begin{table*}[h]
\vspace{3pt} \noindent
\begin{tabular}{p{35pt}p{138pt}p{56pt}p{45pt}p{79pt}}
\hline
\parbox{35pt}{\centering 
\textbf{Scenario}
} & \parbox{138pt}{\centering 
\textbf{Method}
} & \parbox{56pt}{\centering 
\textbf{Success Rate}
} & \parbox{45pt}{\centering 
\textbf{EDP}
} & \parbox{79pt}{\centering 
\textbf{Collision Rate}
} \\
\hline
\parbox{35pt}{\centering 
\textbf{FW}
} & \parbox{138pt}{\centering 
Pure RL Method

Our Method(4 FC layers)
} & \parbox{56pt}{\centering 
0.329

\textbf{0.969}
} & \parbox{45pt}{\centering 
2.313

\textbf{1.177}
} & \parbox{79pt}{\centering 
0.360

\textbf{0.030}
} \\
\hline
\parbox{35pt}{\centering 
\textbf{RO}

\textbf{N=80}
} & \parbox{138pt}{\centering 
Pure RL Method

Our Method(4 FC layers)
} & \parbox{56pt}{\centering 
0.623

\textbf{0.999}
} & \parbox{45pt}{\centering 
1.951

\textbf{0.833}
} & \parbox{79pt}{\centering 
0.268

\textbf{$<$0.001}
} \\
\hline
\parbox{35pt}{\centering 
\textbf{RO}

\textbf{N=120}
} & \parbox{138pt}{\centering 
Pure RL Method

Our Method(4 FC layers)
} & \parbox{56pt}{\centering 
0.432

\textbf{0.960}
} & \parbox{45pt}{\centering 
2.003

\textbf{1.051}
} & \parbox{79pt}{\centering 
0.377

\textbf{0.039}
} \\
\hline
\parbox{35pt}{\centering 
\textbf{CT}
} & \parbox{138pt}{\centering 
Pure RL Method

Our Method(4 FC layers)
} & \parbox{56pt}{\centering 
0.999

\textbf{0.999}
} & \parbox{45pt}{\centering 
1.310

\textbf{0.893}
} & \parbox{79pt}{\centering 
$<$0.001

\textbf{$<$0.001}
} \\
\hline
\end{tabular}
\vspace{2pt}

\caption{The performance of the policy that trained in our basic scenario in different scenarios}
\end{table*}
\end{center}

\section{Conclusion}
In this work, we develop a new framework for the multi-agent pathfinding and
collision avoiding problem by combining a traditional pathfinding algorithm and a
reinforcement learning method together. Both parts are essential--the
reinforcement learning training determines the reliable behavior of the agents,
while the pathfinding algorithm guarantees the arrival of them. \\
Numerical results demonstrate that our approach has a better accuracy and robustness compared to pure learning method. It is also adaptive for
different scale of agents and scenarios without retraining.\\
Ongoing work will concentrate on enhancing the efficiency of our method by
decreasing computational cost using tools in probability theory. Meanwhile, we
are also interested in making use of the cooperation among agents to refine the
framework of current algorithms.

\bibliographystyle{IEEEtran}
\bibliography{ref}

\end{document}